# A Review on the Novelty Measurements of Academic Papers

Yi Zhao, Chengzhi Zhang*

Department of Information Management, Nanjing University of Science and Technology, Nanjing, 210094, China

**Abstract:** Novelty evaluation is vital for the promotion and management of innovation. With the advancement of information techniques and the open data movement, some progress has been made in novelty measurements. Tracking and reviewing novelty measures provides a data-driven way to assess contributions, progress, and emerging directions in the science field. As academic papers serve as the primary medium for the dissemination, validation, and discussion of scientific knowledge, this review aims to offer a systematic analysis of novelty measurements in scientific papers. We begin by comparing the differences between novelty and two similar concepts: originality and innovation. Next, we review the types of scientific novelty. Then, we classify existing novelty measures according to data types and review measures for each type. Subsequently, we survey the approaches employed in validating novelty measures and investigate the current tools and datasets associated with these measures. Finally, we propose a few open issues for future studies.

# Introduction

In the past few decades, scientific knowledge has increased exponentially, fostering major advancements in science. Various measures have been proposed and used to quantify and understand this progress, including scholarly impact (Waltman 2016), interdisciplinarity (Porter & Rafols 2009), novelty (Uzzi et al. 2013), disruptiveness (Park et al. 2023), and so on. Among these, Novelty refers to studies that contribute something new to human knowledge (Shibayama et al. 2021). Scholars argue that novel ideas holds a greater capacity to revolutionize science compared to conventional ideas (Lin et al. 2022). As key mediums for disseminating novel methods, findings, and theories, academic papers draw their vitality and essence from the novelty they present (Cohen 2017; Hou et al. 2022). Therefore, understanding the novelty embedding in academic papers is crucial to driving scientific progress.

    Measuring the novelty of academic papers is one of the pivotal concerns in the realm of scientometrics and research evaluation (Hou et al. 2022). This issue, while of paramount

significance, presents substantial challenges. Novel studies tend to attract more attentions compared to their less novel counterparts, yet they also risk being neglected by the scientific community (Luo et al. 2022). For example, Gregor Mendel's groundbreaking work on the principles of heredity was discovered in 1865. His novel theory, too progressive for his era, initially failed to gain acceptance among his peers. Until 34 years after the novel findings were published, the theory has gradually garnered acknowledgment and recognition (Berry & Browne 2022). The anecdote underscores that the novel research undergoes a protracted journey towards recognition, epitomizing the phenomenon of delayed recognition in novel studies (Raan 2004). This delay can be attributed to the time required for paradigm shifts (Wang et al. 2017). In addition, scientists have to narrowing their expertise to cope with the burgeoning burden of knowledge (Jones 2009). However, the landscape of science is becoming more interdisciplinary (Zhao et al. 2022). The lack of "renaissance men" in modern science compounds the challenge in evaluating the novelty of interdisciplinary research. Collectively, these factors contribute to the complexities inherent in quantifying novelty.

Even though expert-centric approach, such as peer review, is widely endorsed for assessing the novelty of academic papers within the scientific community, it is imperative to acknowledge its limitations (Shibayama et al. 2021). Firstly, the conservative tendencies of peer reviewers often pose challenges in effectively appraising highly novel submissions (Silera et al. 2014). Secondly, the exponential surge in academic publications has not been matched by a commensurate increase in the number of reviewers, potentially leading to a degrade in review quality (Thelwall & Sud 2022). Thirdly, the subjective nature of the review process, influenced by the reviewers' cognition and research experiences, results in varying perceptions of novelty among reviewers (Foster et al. 2021).

With the advancements in information techniques, models from various disciplines have been employed for the quantification of novelty in academic papers, including the BioBERT model (Liu et al. 2022), and the Markov Chain Monte Carlo algorithm (Uzzi et al. 2013). Furthermore, the availability of large-scale bibliographic databases presents an unprecedented opportunity for scientists to objectively and automatically assess the novelty of academic papers. Therefore, our endeavor is to exemplify the wealth of scholarly work by looking through various facets of scientific novelty discussed in the literature. We also review the methodologies designed for quantifying scientific novelty, the assessment of the validity of these proposed approaches, the tools designed for calculating novelty indices, and data with calculated novelty index. Additionally, we also discussed the limitations and potential future direction in the evaluation of novelty in academic papers.

# Definition of scientific novelty, originality, scientific innovation, creativity, and scientific breakthrough

Novelty, as a commonly used concept in the realm of science, constitutes the significant criterion for assessing the academic values and contributions of publications. It is often utilized interchangeably with the terms "originality", "scientific innovation", "creativity", and "scientific breakthrough" (Morgan 1985; Mishra & Torvik 2016). However, there are differences among these concepts.

In the field of scientific publishing, Morgan (1985) believed that novelty connotes that scientific discoveries excite interest and also be new. In an influential work, Uzzi et al. (2013) proposed that novelty is an atypical recombination of elements of prior knowledge from an operational perspective. Packalen and Bhattacharya (2019) illustrated that novelty is the newness of the concepts embedded in the academic papers. Foster et al. (2021) used the definition from the Oxford English Dictionary (OED) to clarify novelty, which characterizes it as "the quality or state of being novel" (Trumble et al. 2007). These definitions share a common characteristic: novelty is assessed in comparison with previous knowledge or combinations thereof, hence bearing a close relationship with temporality and historical context. More specific, the novelty of the scientific papers depends upon the contextual information, including the discipline, the knowledge base used for comparison, and other factors.

As for the concept of originality, there is no universally accepted definition for it. Morgan (1985) defined originality as the state of relating to or constituting a rise or beginning. From the perspective of operationalization, Dirk (1999) conceptualized originality as a recombination of fundamental elements within scientific papers, namely hypotheses, methods and results. This definition bears similarity to the one proposed by Luo et al. (2022) for novelty. In the field of humanities and social sciences, Guetzkow et al. (2004) implicitly delineated originality, believing that the utilization of newly proposed methods, the exploration of previously unexplored data, and the development of new theories all constitute dimensions of originality. This viewpoint implies that originality signifies adding new knowledge to science. In the view of Shibayama and Wang (2019), originality as the extent to which scientific output provides a further study with unique knowledge. This definition highlights the value of focal papers for subsequent studies. Hou et al. (2022) summarize the existing literature on the definition of originality and posited that originality is intrinsically linked to the creation of new knowledge, serving as a catalyst to inspire subsequent research. Thus, different from novelty, originality not only associated with the process of knowledge creation, but also underscores the activating effect of newly created knowledge on the advancement of further knowledge.

Scientific innovation is another pivotal concept, which often appears alongside the term novelty and originality. Schumpeter (1934), often regarded as the founder of innovation studies, first introduced the definition of innovation in his book "*The Theory of Economic Development.*" The author assumed that innovation is the recombination of production factors and conditions within the production system, resulting in the creation of new production function. Furthermore, the term of disruptive innovation represents a specific type of innovation, which is proposed by Christensen in his book entitled "*The Innovator's Dilemma*" (Christensen et al. 2018). This concept is extensively discussed and employed across scientific and technological systems for characterizing technologies or ideas that have the potential to challenge or alter existing technologies and paradigms (Wu et al. 2019; Funk & Owen-Smith 2017). The definitions discussed above draw our attention to three crucial attributes of scientific innovation: first, it embodies new elements or new combinations; second, it confers utility upon individuals, firms, or humankinds; and third, it possesses the ability to advance or disrupt to previous scientific norms and standards.

From a product perspective, creativity refers to the ability to create novel and useful products (Amabile 1983). Similarly, in a scientific context, creativity is the ability to produce novel and useful scientific outputs. According to this definition, we can indicate that creativity can be seen as the driving force behind scientific novelty, while scientific novelty being a key feature of creative scientific outputs.

Most definitions of scientific breakthroughs are quite broad, and there is no universally accepted precise definition of scientific breakthroughs (Wang et al. 2022). Kuhn (1977) argued that scientific breakthroughs are considered as essential paradigmatic shifts in fundamental norms and are associated with scientific revolutions. In Kuhn's opinion, scientific breakthroughs are considered infrequent events. Consequently, some scholars use Nobel-winning publications as proxies for scientific breakthroughs (Min et al. 2021). Wuestman et al. (2020) defined scientific breakthroughs as discoveries that expand the boundaries of knowledge and significantly influence scientific progress, technological advancements, and social development. In the context of science, novelty is typically a fundamental component of a scientific breakthrough; however, novelty is not the only feature, as scientific breakthroughs also emphasize the major impact of the discovery.

Overall, these concepts are closely related, but not equivalent. Novelty emerges by comparing with antecedent elements, and its assessment is inherently context-dependent and somewhat subjective. Originality not only focuses on the creation of new knowledge but also emphasizes its stimulating effect on further studies. Scientific innovation is a broader concept, with novelty and originality serving as its prerequisites (Hou et al. 2022; Janssen et al. 2015). Creativity is associated with human cognitive processes, and novelty is often reflected in scientific outputs as manifestations of creativity for scholars. Scientific breakthroughs are a distinct type of discovery with the transformative power to reshape research directions and scientific paradigms. Novelty is just one of the fundamental characteristics of a scientific breakthrough. In this study, we focus on the novelty of academic papers, which refers to contributing something new to human knowledge and represents the extent to which a contribution differs from existing knowledge. Therefore, the measurement of novelty in academic papers should focus solely on characterizing the relationship between newly generated knowledge and existing knowledge, without the need to trace the flow and impact of the newly generated knowledge.

# Collection of related literature

Firstly, we collected some words about the research object of this paper, encompassing terms such as "novelty", "scientific novelty", "novelty measurement", "combinational novelty", and some words related to research field, including "articles", "academic papers", "scientific research", "academic literature", and so on. Then, we combined these two types of words as search queries and used these queries to search the title/abstract of each paper on Google Scholar, Web of Science (WoS), Springer Link, and ScienceDirect. We manually reviewed the retrieved results to ensure their relevance to the review topic. Using these retrieved papers as seed papers, we also adopted a snowball approach to complement our review by examining the references of the seed papers and the papers citing them. Two specific screening and inclusion criteria were used to construct the final literature collection. Firstly, only the English language articles were included. Secondly, we preserved literature that related to defining, proposing, or validating novelty measures for scientific articles. After the screening, the final corpus comprised a total of 37 publications. The details of these publications are presented in Table A.1 in Appendix A.

# The types of novelty of academic papers

Currently, a widely recognized and unified taxonomy for scientific novelty remains elusive. The philosopher of science, Lakatos (1968), explained the mechanism by which new theories incorporate their antecedents and produce "excess content". The sociologist of science, Mulkay (1974), focused on the new methods, positing that such newly proposed methods are pivotal to the growth of knowledge. Dirk (1999) argued that scientific novelty is comprised of three distinct aspects: new hypotheses, new results, and new methods. In order to better align with the paradigms of social sciences and humanities, Guetzkow et al. (2004) expanded this classification, introducing additional categories such as new theories, new data sources, and new topics. Apart from these theoretical frameworks, consensus in empirical research on the categorization of scientific novelty has yet to be established. The categorization of scientific novelty reviewed herein is summarized in Table 1. Considering the different research objectives and the classification standards used in those literature, we have divided the topology of novelty into two main categories: conceptual approach and the degree of novelty.

  The first class of studies utilizes several theories to conceptualize measures for assessing scientific novelty. These theories facilitate the classification of novelty measures. For example, Mishra and Torvik (2016) believed that essence of scientific novelty is linked to the age and volume of concepts embedded in the papers, and novel papers feature rare and newly-emerged concepts. This conceptual approach enables the authors to classify scientific novelty into two types: Time Novelty and Volume Novelty. Yan et al. (2019) summarized prior studies on quantifying scientific novelty. They argued that the scientific novelty predominantly steams either from new combinations of existing knowledge or the creation of new knowledge elements. Consequently, they suggested that scientific novelty consists of two dimensions: new combinations and new components. In another study, Luo et al. (2022) embraced combinatorial innovation theory and life-cycle theory to design measures for quantifying scientific novelty. Drawing on terms of "question" and "method" as proxies for knowledge elements, they designed two novelty metrics centered around the combinations of these elements and their "academic age". They believed that these measures, namely Life-Index Novelty and Combinational Semantic Novelty, encapsulate two facets of scientific novelty. In addition, Shi and Evans (2023) indicated that scientific novelty has two foundational dimensions: content and context. The authors argued that a new article comprising contents, such as questions, concepts, and methods, which have not been previously co-appeared, could be deem more novel. Likewise, the old content presented in new contexts, including conferences and journals, may also attain a veneer of novelty for new audiences. Consequently, the authors classified novelty into two categories: Content Novelty and Context Novelty. Based on Dirk (1999) topological framework of novelty, Leahey et al. (2023) delineated three type of novelty: new results, new theories, and new methods. Overall, these studies focus solely on the varied types of scientific novelty, and refrain from quantifying the degree of novelty.

  The second class of literature generally classifies scientific novelty into different categories based on the degree of novelty. These scholarly works initially design metrics to quantify the novelty of academic papers. Subsequently, those academic papers are assigned into different categories based on their novelty scores. For example, Uzzi et al. (2013) introduced a z-score method for measuring novelty, where z-score values determine the classification of a paper as either novel or conventional. Additionally, other classification systems are relatively simple (Wang et al. 2017; Ke 2020). For instance, Trapido (2015) divided novelty into Low and High Novelty.

Determining the optimal classification system for novelty of academic papers remains difficult due to the variability in measure design and the diverse fields to which these measures apply. This challenge is compounded by the unresolved question of whether each measure captures unique dimensions of novelty or merely reflects similar ones (Foster et al. 2021). Moreover, due to the disparity in research paradigms across disciplines, developing a universally recognized classification system for novelty is challenging.

Table 1 The types of scientific novelty

| Category | Types of novelty | Knowledge proxy | Domain | Typical literature |
|---|---|---|---|---|
| Conceptual approach | Time Novelty, Volume Novelty | MeSH terms[1] | Life sciences | Mishra and Torvik (2016) |
| | Recombination Novelty, New Elements Novelty | keywords | Wind energy | Yan et al. (2019) |
| | Life Index Novelty, Combinational Semantic Novelty | Research question terms and research method terms | Computer science | Luo et al. (2022) |
| | Content Novelty, Context Novelty | Mesh terms, journals, conferences | Life sciences, physical sciences | Shi and Evans (2023) |
| | New Results, New Theory, and New Methods | | Physical sciences, Technology, Social sciences, Life sciences | Leahey et al. (2023) |
| | Novel Combinations, Conventional Combinations | Journals | Science and engineering, Social science, Arts and humanities | Uzzi et al. (2013) |

---

[1]MeSH (Medical Subject Headings) is a set of controlled vocabulary terms that is hierarchically structured and developed by the National Library of Medicine.

| Degree of novelty | Non-novel, Moderately Novel, Highly Novel | Journals | All disciplines | Wang et al. (2017) |
| --- | --- | --- | --- | --- |
| | High Novelty, Low Novelty | Journals | Information theory | Trapido (2015) |
| | Non-novel, Moderately Novel, Highly Novel | MeSH terms | Biomedical science | Ke (2020) |

# The conceptualization of scientific novelty

Numerous methods have been proposed to measure scientific novelty, which can be divided into four groups based on their conceptual foundation, as illustrated in Figure 1. Some scholars believed that scientific novelty consists of two dimension: new elements and new combinations (Yan et al. 2019; Luo et al. 2022). The former highlight the uniqueness of the newly proposed knowledge elements, also referred to as "Absolute Novelty" by Berlyne (1960). For instance, the SARS-CoV-2 virus, first discovered in 2019 (Zhu et al. 2020), is unique in many ways in human history and can be considered a novel discovery for the scientific community from the perspective of science studies. The latter dimension focus on the recombination of pre-existing knowledge elements. This perspective stems from the idea that scientific novelty arises from the recombination of theories, methodologies, question, data and topics (Guetzkow et al. 2004; Nelson & Winter 1982), inspired by the Schumpeter (1934) recombination theory. For example, the birth of the first usable computer mouse is attributed to the combination of electronics and the trackball (Hargadon & Sutton 1997).

Some scholars argue that the quantification of scientific novelty can be conceptualized through a network perspective (Foster et al. 2015; Hofstra et al. 2020). These novelty measures represent the scientific paper as networks, comprising knowledge elements and links among them. From a macro point of view, the tightly connected components can be regarded as a "knowledge community". Novelty, in this context, is ascribed to new connections that disrupt existing network structures, i.e., those that bridge two distinct communities. Moreover, a limited yet significant corpus of literature has introduced the concept of *surprise* from the field of psychology to conceptualize novelty measures. These studies consider the violation of expectations as novelty (Foster et al. 2021). The extent of surprise a scholar experiences by a scholar when reading an academic paper is directly proportional to its perceived novelty. From this perspective, the novelty of a paper depends on the cognitive level of readers, implying that a paper may contribute novel knowledge to one individual while appearing less novel to another.

Taken together, from a broad perspective, the novelty is determined by the extent to which an idea differs from current knowledge and templates. The evaluation of novelty is inherently influenced by the scope and depth of the knowledge repository utilized in the comparison. It is important recognize that a universal knowledge base, encompassing every discipline and temporal phase, does not exist. Therefore, almost every measure is limited to gauging the localized novelty of a paper within a given context.

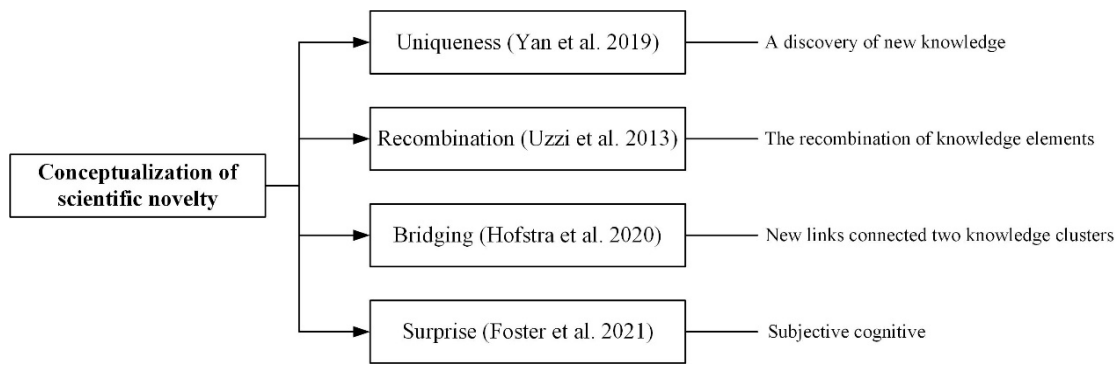

Figure 1 the conceptualization of scientific novelty

# Methods for measuring novelty of academic papers

Researchers have conducted various studies to evaluate the novelty of scientific papers. Due to the complexity of novelty evaluation and disciplinary heterogeneity, there is currently no generally recognized method that can be applied to different disciplines to evaluate the novelty of scientific papers, except for peer-review. The aim of this paper is to provide a comprehensive overview of insights into the quantitative evaluation methods of scientific novelty; therefore, we will not discuss qualitative evaluation methods extensively in this review. This section will review the three main types of novelty measures shown in Figure 2. From the operationalization perspective, novelty measures can be roughly categorized into three groups: citation-relations based measures, textual-data-based measures and measures that leverage multitype data.

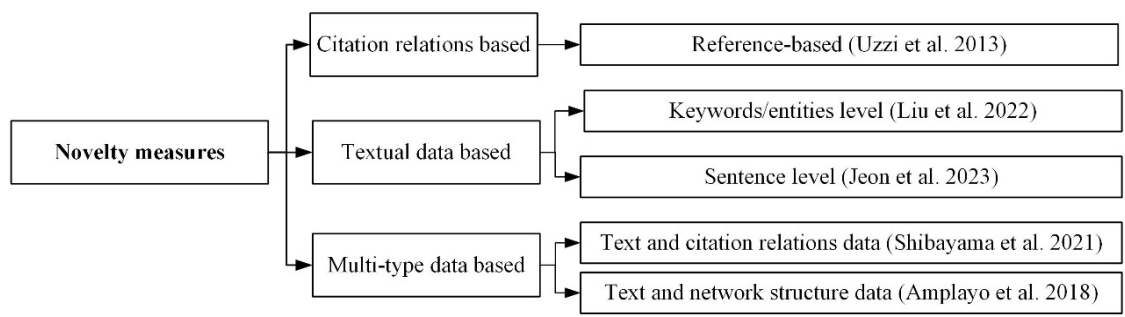

Figure 2 three main types of novelty measures

## Novelty measures based on citation relations

The citation relations are commonly employed in the development of indicators aimed at quantifying novelty of academic papers. Considering the diverse structures of citation relations, the existing measures of novelty can be categorized into two distinct groups: those based on references.

From a conceptual standpoint, reference-based measures emphasize the recombination of pre-existing knowledge elements. Here, a novel combination is identified when there is a distant or unusual combination of these knowledge elements, termed 'Relative Novelty' (Berlyne 1960). In these measures, journals are considered knowledge elements, as journal-level analysis is optimally suited to distinguish between different fields of knowledge (Small 1973). For example, an early noteworthy study by Uzzi et al. (2013), the authors introduced the Z-score measure for quantify novelty, focusing on the atypical combination of referenced journals. Specifically, their approach to measuring novelty involved three steps. First, journal pairs were formed through the co-citation of referenced papers. Second, a Null model was applied to normalize the number of journal pairs, with the resulting value serving as a novelty score for journal pairs, namely z-score. The final step involved evaluating the cumulative distribution of the z-score and the novelty of each article determined by its tenth percentile. However, this method is costly in terms of computing resources and time. Later on, Lee et al. (2015) built upon Uzzi et al. (2013)'s idea, modifying the calculation process. Their method involved two stages. First, the commonness of reference journal pairs was computed using the entire WoS database. This involved collecting all journal pairs from articles published in the same year to construct a journal pair pool for each year. The commonness of journal pair $i$ and $j$ in year $t$, denote as $commonness_{ijt}$, was calculated using the ratio of the observed to expected frequency of these pairs. The second stage involved calculating a paper's commonness for a given paper $t$ by recording the $commonness_{ijt}$ values of all journal pairs cited in that paper, with the $10^{th}$ percentile representing the paper's commonness. The paper's novelty was then defined as the inverse of its commonness. Further improvements to this novelty measurement were proposed by Wang et al. (2017). Their approach sought to examine the novelty of scientific papers by considering both the newness of reference journal pairs and ease of forming new journal combinations. A journal pair was considered new if it had not appeared in previous publications. The ease of forming new combinations was assessed by examining the similarity of co-citation profiles of the paired journals. Papers lacking new journal pairs were categorized as non-novel. This method greatly optimizes the computation complexity of novelty measurement proposed by Uzzi et al. (2013).

Research papers have also been utilized to measure the novelty of scientific papers. For example, Trapido (2015) formulated a novelty index to encapsulate the unusualness of knowledge recombination in scholarly papers that refer back to antecedent knowledge. This measure specifically pairs the focal paper $i$ with former papers $j$ that have been published by authors who are in the same field as focal paper $i$. The overlap score $os_{ij}$ is calculated as the number of publications cited by both $i$ and $j$, then dividing it by the summation of unique citations in $i$ or $j$. Subsequently, the average $os_{ij}$ is interpreted as usualness of knowledge recombination. Finally, the average $os_{ij}$ subtracts maximum value within a specific year is used to convert the usualness to unusualness.

## Novelty measures based on textual data

Textual information in a scientific article, such as the title, abstract, and body text, is the important part of scientific article, which is also utilized to quantify the novelty of scientific articles. Generally, novelty measures for scientific articles can be implemented according to different levels of granularity of textual data, including keywords, entities, and sentences.

**(1) Keywords-based measures**

Some novelty measures are constructed based on the keywords or knowledge entities[2]. if an article contains words or combinations of words that are new to the discipline, an article may be deemed to contribute novel ideas to its research discipline if it contains words or combinations of words that are new to the field (as shown in Table 2).

**Table 2 Keywords-based measures of novelty**

| Authors | Knowledge proxy | Domain | Formulation |
|---|---|---|---|
| Azoulay et al. (2011) | Mesh term | Life sciences | $Novelty = \frac{\sum_1^n \text{MeSH term age}}{n}$ |
| Mishra and Torvik (2016) | Mesh term | Life sciences | Time Novelty: the number of years since the first appearance of a concept. Volume Novelty: the number of articles since the first appearance of a concept |
| Yan et al. (2019) | Keywords | Wind energy | $New\ Combinations = \frac{new\ pairs\ of\ keywords}{all\ potemtial\ pairs\ of\ keywords}$ $New\ Components = \frac{new\ appearing\ keywords}{all\ potemtial\ pairs\ of\ keywords}$ |
| Ruan et al. (2023) | Mesh term | Biomedical and life science | $Term\ Novelty_{a,b} = -\ln \frac{obs(a,b)}{\exp(a,b)}$ $Paper\ Novelty = \frac{\sum_1^n \text{Novelty of all term pairs}}{n}$ |

For instance, using MeSH terms, Azoulay et al. (2011) proposed that the average age of these terms could serve as an indicator of an academic paper's novelty. The term's age refers to number of years since its initial appearance in the PubMed database. In the formula, *n* represents the number of MeSH terms in a paper (see Table 2, first row of the Formulation column). Mishra and Torvik (2016), expanded on this, suggesting that a paper's novelty could be encapsulated by the age and frequency of use of its MeSH terms. From this premise, the authors introduced two categories of novelty: Time-based and Volume-based Novelty. They took the youngest MeSH terms and the youngest pairs of MeSH terms to denote a paper's novelty. Ruan et al. (2023) recently posited in their work that MeSH-terms are a better proxy of knowledge when gauging novelty than references. Consequently, they proposed a MeSH-term-based novelty index, building upon the study carried out by Uzzi et al. (2013). This index utilizes the ratio of observed and expected frequency of MeSH term pairs to determine their commonness. The novelty of the topic is then defined as the logarithmic transformation of commonness with a negative sign. The paper's novelty is calculated as the average novelty of all the MeSH pairs. In the formula, *obs*(*a, b*) and *exp*(*a, b*) denote the observed and expected frequencies of the MeSH pair *a* and *b*, while *n* represents the number of potential MeSH pairs in a focal paper (see Table 2, fourth row of the Formulation column). However, it's important to note that the application of MeSH terms is primarily restricted to one discipline: biomedical and life science. To address this limitation, Yan et al. (2019) employed keywords as a proxy for knowledge components in scientific papers. Following the work by Verhoeven et al. (2016), the authors argued that a paper's novelty has two dimensions: new components and new combinations. New components refer to the keywords that haven't appeared within a research discipline in the previous five years, and the paper's novelty can be determined as the ratio of these new keywords to the total number of keywords in the paper. New combinations are defined as the unique keyword pairs that have never appeared within a research field in the preceding five years. The paper's novelty, with respect to new combinations, is calculated as the percentage of these unique pairs to all possible keyword combinations in the paper. However, this approach is also not devoid of limitations. It can be challenging to determine whether keywords represent the paper's topic or highlight specific novel contributions. Furthermore, the functionality of the keywords is frequently overlooked in these novelty indexes.

---

[2] Although keywords and entities are different concepts, they are all words or short word sequence in a broad view.

**(2) Entities-based measures**

All of the aforementioned novelty metrics ignore the semantic relationships among keywords, which might introduce bias into novelty measurements. For example, the phrase "neural network" usually refers to a bionic algorithm, or more specifically, an artificial neural network, in the field of computer science. However, in biology, it denotes a complex network comprising neurons and their connections.

As illustrated in Table 3, Liu et al. (2022) proposed an entity-based novelty measure and takes the semantic relationships among entities into account. The authors first created pairs of all bio-entities extracted from CORD-19 papers, calculating the cosine distance between two entities in each pair. Then, they sorted these entity pairs in descending order based on the cosine distance. They determined a paper's novelty by the ratio of novel bio-entity pairs to the potential number of such pairs that the paper could contain, considering those in the top 10 percent of the cosine distance distribution as novel pairs. Nevertheless, this method fails to account for the temporal aspect of entities and doesn't differentiate between types of entity pairs. To address this issue, a recent study by Luo et al. (2022) developed a new approach for novelty quantification from the perspective of question-method combinations. They suggested two novelty indicators encompassing temporal, frequency, and semantic aspects: life-index novelty and semantic novelty. They extracted all question and method entities from computer science abstracts, recording the "birth year" of all entities, their combinations, and their accumulated frequency. The life-index novelty of a paper is the average of the novelty score of the questions, the methods, and the question-method combinations in the paper. The novelty of a question is measured by multiplying the age of the question entity by its frequency. The computation for other forms (i.e., method novelty, question-method novelty) of novelty follows the same trajectory. Semantic novelty comprises the novelty of questions, methods, as well as question-method combinations. Semantic novelty of a question-entity is measured by one minus the highest cosine similarity between the current question-entity and the existing ones in the corpus. A method-entity's semantic novelty is calculated similarly. Regarding the question-method combinational novelty, if the question is new and the method is old, the novelty is dictated by the semantic novelty of the question. This is accomplished by comparing its semantic similarity to questions previously associated with the older method. In the formula, *q* and *m* represent the question and method entities, respectively (see Table 3, second row of the Formulation column). This method offers a fresh viewpoint for researchers to quantify novelty. However, its generalizability is restricted as not all disciplines adhere to the same research paradigm. Chen et al.

(2024) used four types of entities extracted from full text as proxies of knowledge, including methods, data sets, tools and metric. Similar to the idea of Liu et al. (2022), the ratio of novel entities pairs to the number of potential entity pairs in a focal paper are used to calculate novelty score. In the formula, *m* represents the number of novel entity pairs, while $C_n^2$ denotes the number of potential pairs in a paper (see Table 3, third row of the Formulation column). Interestingly, the authors also analyze the contributions of different types of entity combinations to the novelty score of academic papers. The results show that method-related combinations, such as method-method pairs, method-dataset pairs, and method-tool pairs, contribute more to the novelty of academic papers compared to other entity combinations.

**Table 3 Entities-based measures of novelty**

| Authors | Knowledge proxy | Knowledge extraction Source | Domain | Formulation |
|---|---|---|---|---|
| Liu et al. (2022) | Bio-entities: species, disease, gene and chemical. | Abstract | Coronaviruses | $Novelty\ Score = \frac{number\ of\ novel\ entity\ pairs}{all\ potemtial\ entity\ pairs}$ |
| Luo et al. (2022) | Questions and methods | Abstract | Computer science | $Life\ Index\ Novelty = \frac{1}{3}(LifeIndexScore(q) + LifeIndexScore(m) + LifeIndexScore(q,m))$ $Sematic\ Novelty = \frac{1}{3}(SematicNovelty(q) + SematicNovelty(m) + SematicNovelty(q,m))$ |
| Chen et al. (2024) | Knowledge entities: methods, data sets, tools and metric | Full text | Natural language processing | $Novelty\ Score = \frac{m}{C_n^2}$ |

**(3) Sentences-based measures**

Few scholars have utilized the sentence-level characters to measure novelty, as shown in Table 4.

**Table 4 Sentences-based measures of novelty**

| Authors | Knowledge proxy | Knowledge extraction Source | Domain | Formulation |
|---|---|---|---|---|
| Chen and Fang (2019) | N-grams | Abstract | Semantic analysis | $Novelty = \begin{cases} 1, \alpha(a) \cap \alpha(b) \neq \emptyset \\ 0, \alpha(a) \cap \alpha(b) = \emptyset \end{cases}$ |

| Jeon et al. (2023) | Title | Title | Biomedical science | $Novelty\ score_i = \text{LocalOutlierFactor}(d_i)$ |
| Wang et al. (2024) | Contribution sentence | Full text | Computer science | $Novelty(Paper_p) = \sum_{i=1}^{n} \theta_i \times Novelty(C_i) + \varepsilon$ |

Chen and Fang (2019) developed an automatic method for extracting novel ideas. They analyzed the abstracts of papers published between 2001 and 2017 in the field of semantic analysis and extracted N-grams as potential novel ideas. After eliminating any irrelevant N-grams, the remaining ones were cross-checked via the Scopus database API to determine their novelty. In the formula, $\alpha(a)$ represents N-grams extracted from focal papers, while $\alpha(b)$ denotes N-grams in the Scopus database (see Table 4, first row of the Formulation column). Despite its direct approach to evaluating novelty in scientific papers, this method fails to qualify the degree of novelty. Novelty detection, that is, determining the newness of elements compared to previous related documents, offers insights for scientific measurements. To capture implied semantic knowledge, Jeon et al. (2023) proposed a method, combining fastText and local outlier factor approach, to measure the novelty of scientific articles. In this approach, fastText is employed to embed each article's title into a vector space, and the local outlier factor is used to quantify the extent to which the paper is isolated from its neighbors. Nevertheless, this method can only capture the local novelty of scientific papers. Wang et al. (2024) believed that research contributions describe how a current study adds new knowledge in contrast to existing studies on a specific topic and can be used to measure the novelty of scientific papers. Therefore, the authors proposed a method for measuring the novelty of academic papers, which includes three steps: first, they used TextCNN and transformer-based models to identify the contributions sentences; second, they integrated the BERTopic and cloud models to represent each contribution sentences; finally, they applied the cloud similarity to measure the novelty of the scientific papers. In the formula, $Novelty(C_i)$ represents the novelty of the *i-th* type of contribution cloud $C_i$, whereas $\theta_i$ denotes the weight of the *i-th* type of contribution in the paper *p*. (see Table 4, third row of the Formulation column).

## Novelty measures based on multiple types of data

Several novelty measures have been developed that adeptly incorporate both textual information and citation relation data. Shibayama et al. (2021) argued that the novelty of a scientific paper can be determined by the semantic distance of between its references. Based on this assertion, they begin by converting the titles of every cited reference in a focal paper into vectors. The mean vector of the word embeddings of all words in the title is used to represent the vector representation of the title. Next, they calculate the cosine distance of all pairs of cited references within this focal paper. The final step involves adopting the q-percentile of these cosine distances score as a proxy for the paper's novelty.

The aforementioned measures of novelty are largely predicated upon the theory of recombinant innovation. In contrast, some scholars have approached the concept of scientific novelty through the lens of network analysis. Foster et al. (2021) have termed network-based measures as "structural measures". Amplayo et al. (2018) postulated that a paper connecting previously unconnected networks of knowledge should be deemed more novel; they thus introduced a network-based approach to evaluate the novelty of scholarly articles. This method entails the initial creation of varied entity-based graphs, with nodes ranging from the micro (e.g., keywords, words, topics) to the macro scale (e.g., authors, documents). These graphs are constructed using entities derived from publications within a specific time frame and serve as foundational "seed graphs." Subsequently, papers published after this period are integrated into these seed graphs one at a time, altering their characteristics. The degree of change within the graphs furnishes features to train an autoencoder neural network. The error rate of the training is interpreted as an indicator of novelty; a higher error rate means that replication of the data requires more effort, signifying that the knowledge contained in the paper is novel. In 2020, Hofstra et al. (2020) developed a structural approach to measure the novelty of PhD theses. This method utilized the scientific concepts extracted from the abstract of theses to construct a conceptual network. In this network, nodes present concepts, and the edges are formed when two concepts co-occur within an abstract. A thesis that is the first to connect two concepts establishes what is considered a new link. The number of such new links within a thesis is then viewed as an indicator of the thesis's novelty. However, the deficiency of this approach lies in its absence of a universally recognized definition and methodology for concept extraction. Martín de Diego et al. (2021) focused on evaluating the novelty of medical research papers. The basic idea behind this novelty metric is to quantify the disparity between current knowledge graphs and antecedent knowledge bases. To this end, the authors extracted medical concepts from articles and built a text graph based on these extracted concepts, with nodes representing the medical concepts and edges linking concepts featured within the same article. The novelty measure integrates two distinct components: the number of overlapping concepts between disease subgraph and current text graph and the relevance of their common relationships. Owing to the voluminous nature of conceptual studies in medical research, the concepts are more easily defined and extracted than in other fields. This is the key strength of using concepts (i.e., MeSH terms) to study novelty quantification in medical research.

# The validation of the proposed novelty measures

Despite the academic community have proposed numerous methods for measuring the novelty of scientific paper, the validity and quality of these methods remain of significant concern. To ascertain whether the proposed measures accurately capture the true novelty of scientific work, two types of validation methods are commonly employed: direct and indirect.

The direct method primarily examines the correlation between the proposed novelty metrics and the ground truth, such as evaluations by experts or results from peer reviews. For example, Tahamtan and Bornmann (2018) identified landmark papers in scientometrics, and conducted interviews with the corresponding authors to examine the role that cited references played in the genesis of ideas for their papers. The authors summarized the responses and concluded that cited references have minimal impact on idea generation and, therefore, are not relevant to the measurement of scientific novelty. However, the limited number of interview participants restricts the generalization of these conclusions. In another study, Bornmann et al. (2019) compared novelty measures with assessments of novelty conducted by experts in order to validate the quality of novelty measures conceptualized from the perspective of reference recombination. Specifically, they collected the peer assessment results for the published articles from F1000Prime platform, and compared it with Uzzi et al. (2013)'s novelty measure (i.e., U score) and Wang et al. (2017)'s novelty measure (i.e., W score). The results suggested that the U score aligns with peer evaluations, whereas the W score does not. In related research, Shibayama et al. (2021) conducted surveys wherein scientists were prompted to assess their selected articles in four categories: theory, phenomenon, method, and material. The researchers then determined the correlation between these proposed novelty measures and the self-reported novelty scores. The outcomes revealed a significant correlation between the measures and self-reported innovation specifically in theory, material, and phenomenon, affirming the measures' efficacy. Additionally, Matsumoto et al. (2021) employed a similar method to evaluate the effectiveness of a newly proposed novelty index. These methodologies serve well for validating novelty measures, yet the significant costs associated with gathering peer reviews on a large scale hinder the extensive adoption of such direct assessment methods. Additionally, Fontana et al. (2020) adopted four methods to test the novelty measures proposed by Uzzi et al. (2013) and Wang et al. (2017). The results show that Uzzi et al. (2013)'s measure overlaps with interdisciplinarity, while Wang et al. (2017)'s measure only captures the structural features of citation networks.

Indirect methods are relatively straightforward and cost-effective to scale up, including calculating the correlation between the current novelty index and previously proposed novelty indices, or using the current novelty index as a predictor of citation counts. Several scholars have adopted the correlation tests to confirm the consistency of proposed novelty methods with earlier ones (Luo et al. 2022). The efficacy of this method presupposes that the proposed measures of novelty are capable of capturing the true novelty of scientific papers. Nevertheless, this precondition does not always hold true (Foster et al. 2021). Other scholars have used novelty as a predictor of scientific impact, assessing the prediction performance to gauge the effectiveness of the novelty index (Foster et al. 2021; Shi & Evans 2023; Luo et al. 2022). However, this method also has limitations. The relationship between novelty and impact is often assumed to be linear (Lee et al. 2015), but some studies suggest it follows an inverted-U-shaped relationship (Yan et al. 2019). Thus, this relationship remains equivocal, which undermines the reliability of such evaluative methods. Moreover, citing practices are complex, driven by a variety of factors (Kaplan 1965). In such cases, articles may be cited arbitrarily and bear no relevance to their novelty.

# Existing novelty calculation tools and data with calculated novelty index

Open science resources, such as data, tools, and codes, has become essential factors driving the rapid development of the scientific community (Mirowski 2018). In the science of science field, many scholars have developed several tools for computation of the novelty index. Additionally, scholars have also provided open access to data that already contains a precomputed novelty index.

There are two publicly available tools for novelty calculation (In Table 5). The first tool, *novelpy*, was released in 2022 (Pelletier; & Wirtz; 2022). This open-source Python package supports the calculation of most well-known novelty and disruptiveness indices, including atypicality (Uzzi et al. 2013), commonness (Lee et al. 2015), bridging (Foster et al. 2015), scientific novelty (Wang et al. 2017; Shibayama et al. 2021), breadth/depth (Bu et al. 2021) an disruption (Wu et al. 2019; Funk & Owen-Smith 2017). The second tool, *pySciSci* (Gates & Barabási 2023), is designed for the analysis of large-scale bibliometric data. It is also implemented in Python programming language. In contrast to *novelpy*, *pySciSci* supports not only the computation of novelty-related indicators, such as atypicality (Uzzi et al. 2013) and disruption (Wu et al. 2019) but also other commonly used indicators, such as Sleeping Beauty score (Ke et al. 2015), Q-factor (Sinatra et al. 2016), and so on.

**Table 5 The compassion of two tools for novelty calculation**

| Tool name / Type | Novelpy | pySciSci |
|---|---|---|
| Programming language | Python 3.x | Python 3.x |
| Supported indicators | atypicality (Uzzi et al. 2013), commonness (Lee et al. 2015), bridging (Foster et al. 2015), scientific novelty (Wang et al. 2017; Shibayama et al. 2021), breadth/depth (Bu et al. 2021) an disruption (Wu et al. 2019; Funk & Owen-Smith 2017) | atypicality (Uzzi et al. 2013), disruption (Wu et al. 2019), Sleeping Beauty score (Ke et al. 2015), Q-factor (Sinatra et al. 2016), and so on. |
| Function of tools | Calculation for novelty indicators | Data preprocessing, indictors calculation, network analysis |
| Supported dataset | MAG, WOS, APS, PubMed, DBLP, and OpenAlex | Data must be converted to JSON or MongoDB format |

Beyond the mentioned tools, and to the best of our knowledge, there is currently one freely available dataset with precomputed novelty indicators. To reduce the barrier to entry and enhance the ability to replicate new findings, Lin et al. (2023) introduce a large-scale open dataset known as *SciSciNet*. This dataset draws on the Microsoft Academic Graph (MAG) and integrate multiple datasets, spanning from upstream funding sources to downstream public uses. In addition, this dataset includes the calculation of several commonly used indicators, including not only atypicality (Uzzi et al. 2013), but also other indicators such as team size, disruption (Wu et al. 2019) and more.

In general, these resources share a similar goal: to aid researchers in gaining a better understanding, simplify usage, and choose the appropriate novelty measures for their innovative research. However, differences exist in terms of these two novelty calculation tools. *Novelpy* focus on facilitating the calculation of novelty-related indicators, whereas *pySciSci* is primarily designed for scientometrics and bibliometrics studies, covering the entire process from data loading to analysis. Furthermore, data must be converted to JSON or MongoDB format before using *Novelpy*, while *pySciSci* is compatible with several major datasets, such as MAG[3], WOS[4], the American Physics Society (APS)[5], PubMed[6], the DBLP Computer Science Bibliography (DBLP)[7], and OpenAlex[8].

# Summary and outlook

By reviewing relevant literature, we find that a series of novelty measures have been proposed. Scientific novelty is conceptualized as the emergence of newly discovered knowledge elements or an unusual combination of existing knowledge elements. Furthermore, novelty in science can also be assessed by whether a study produces new knowledge that unites previously independent knowledge communities. From an operational standpoint, various factors are used as proxies for scientific knowledge, including cited articles, referenced journals, MeSH terms, entities, and titles. These attributes can generally be categorized into two groups: citation-relations data and textual data. In addition, natural language processing techniques are gaining popularity in the development of novelty indicators. Despite some progress, there is ample space for enhancing the efficacy of novelty measurement. In addition, novelty indicators play a vital role in the measurement of scientific progress.

## Constructing a comprehensive ground-truth corpus for the evaluation of novelty measures

Although many novelty measures have been proposed, the effectiveness of these measures remains a matter of debate due to the absence of a unified, comprehensive ground-truth corpus. To our knowledge, the underlying cause is the lack of a unified comprehensive ground-truth corpus. To our knowledge, various existing corpora, including F1000Prime data (Bornmann et al. 2019), questionnaire surveys (Shibayama et al. 2021), and Nobel prize-winning papers (Shi & Evans 2023), have been utilized to validate proposed novelty measures. However, these corpora have several limitations. Post-publication evaluation on F1000Prime platform is exclusive to the biomedical field, and such evaluations might be biased due to the subsequent influence following publication, and therefore might not reflect pure novelty as defined in our study. Nobel prize-winning papers report revolutionary findings. Therefore, these papers not only emphasize the novelty of the research but also the uptake, reuse, and real-world impact of the findings. Furthermore, the size of the Nobel prize-winning papers corpus is notably small. To overcome these limitations, future research should consider constructing a comprehensive ground-truth corpus for novelty measure evaluation. While this endeavor may be time-consuming and expensive, it is of significant importance. Another approach would be to incorporate open pre-publication reviews into the ground-truth corpus. There are several open review platforms and journals that share pre-publication reviews, including Publons[9], Openreview[10], Nature communication[11], eLife[12] and so on. More specifically, the International Conference on Learning Representations, using the OpenReview mechanism, requires reviewers to provide scores on technical/empirical novelty of papers in some years. These scores could serve as the gold standard for measuring the novelty of academic papers. In addition, research contributions sentences can be viewed as the author's summary of how their work contributes new insights to existing knowledge. Therefore, research contributions sentences could also serve as a ground-truth corpus for evaluating novelty measures.

---

[3] https://www.microsoft.com/en-us/research/project/microsoft-academic-graph/
[4] http://webofknowledge.com
[5] https://www.aps.org/
[6] https://pubmed.ncbi.nlm.nih.gov/
[7] https://dblp.org/
[8] https://openalex.org/

## Intensifying theoretical research on scientific novelty

Several studies have argued that novelty features multiple dimensions or types, but existing novelty measures only capture a single aspect (Yan et al. 2019; Leahey et al. 2023). However, the scientific community has yet to reach a consensus on the dimensions of scientific novelty. Scholars from social sciences and humanities believed that scientific studies have six types of novelty, including new hypothesis, new result, new method, new theory, new data, and new topic (Guetzkow et al. 2004). From the perspective of operationalization, Shi and Evans (2023) suggested that scientific novelty consists of two aspects: content and context novelty. Moreover, current novelty measures are heterogenous. It remains uncertain whether all novelty measures capture the same aspect or distinct, yet undefined aspects of novelty. One potential way to fill this gap is to put more effort into theoretical research on scientific novelty. Before proposing measures for scientific novelty, scholars need to understand where scientific novelty comes from and what aspects constitute it. Understanding these theoretical bases could help researchers verify whether these indicators accurately measure what they intend to measure.

## Implementing automatic novelty assessment using multi-source data and advanced machine learning techniques

Most novelty measures are employed using either citation-related data or textual data. Among these measures, those based on textual data largely exploit entity-level and sentence-level data. Leveraging the open-access full-text data and advanced text mining techniques, particularly the adoption of large language models, may provide a pathway for automated novelty evaluation. Thelwall (2024) argued that ChatGPT is not accurate enough to directly perform research evaluations for academic articles. However, large language models like ChatGPT can be used to improve the extraction of knowledge entities, contribution sentences and other content related to scientific novelty. Indirectly, this can enhance the accuracy of novelty measures.

---

[9] https://publons.com/
[10] https://openreview.net/
[11] https://www.nature.com/ncomms/
[12] https://elifesciences.org/

Moreover, by using effective prompt engineering strategies, large language models can provide useful feedback on the novelty of an academic paper, representing the knowledge of such models. By integrating the knowledge of language models, citation-related data and textual data, we can develop models for novelty predication.

## Proposing a multi-dimensional framework for measuring scientific advancements

Scientific progress is driven by significant, breakthrough discoveries, and novel research is more likely to lead to scientific breakthrough (Rodríguez‐Navarro 2015; Veugelers & Wang 2019). Scientific novelty can be viewed as the production of new knowledge for humankind. Therefore, measuring scientific novelty helps stakeholders understand scientific advancements from the angle of how research extends and enriches existing knowledge. Since the progress of science is pretty complex, simply relying on scientific novelty to measure scientific progress is far from enough. In addition to novelty, many indicators can be used to measure the advancement of science, including the number of highly cited papers, number of Nobel Prize-winning discoveries, interdisciplinarity, disruption index, and so on (Funk & Owen-Smith 2017; Rodríguez‐Navarro 2015). In the future, it is highly necessary to propose a multi-dimensional framework for measuring scientific advancements. Experts from different disciplines could form a committee with the aim of developing clear, objective criteria for evaluating scientific progress across varied fields. Moreover, while this study focuses on the measures for assessing novelty in academic papers, future research could explore the transferability of these evaluation metrics to other types of scientific outputs. This could also aid the more comprehensive and accurate evaluation and understanding of scientific progress.

### The potential implications of scientific novelty measures

The novelty measurement of academic papers has become increasingly important in the scientific community. For example, given the huge volume of scientific papers, the automatic novelty measures can help editors and reviewers quickly evaluate an article's degree of novelty. Automatic novelty measures have the potential to enhance the efficiency and accuracy of the peer review process. Additionally, novelty measures are also beneficial for funding agencies. Typically, number of publications and citation counts are used to assess the research capabilities of scholars. However, these metrics can be biased. Novelty measures can act as a complementary metrics for scholar evaluation, helping to mitigate biases inherent in metadata-based evaluation methods.

# Acknowledgments

This study was supported by the National Natural Science Foundation of China (Grant No. 72074113) and the Open Funding Project of Laboratory of ISTIC-Springer Nature Joint Laboratory for Open Science (Grant No. ISN23012).

# Declarations

# References

Amabile, T. M. (1983). The social psychology of creativity: A componential conceptualization. *Journal of Personality and Social Psychology, 45*(2), 357-376.

Amplayo, R. K., Hong, S., & Song, M. (2018). Network-based approach to detect novelty of scholarly literature. *Information Sciences, 422*, 542-557. https://doi.org/10.1016/j.ins.2017.09.037

Azoulay, P., Graff Zivin, J. S., & Manso, G. (2011). Incentives and creativity: evidence from the academic life sciences. *The RAND Journal of Economics, 42*(3), 527-554. https://doi.org/10.1111/j.1756-2171.2011.00140.x

Berlyne, D. E. (1960). *Conflict, arousal, and curiosity* (Conflict, arousal, and curiosity.). New York, NY, US: McGraw-Hill Book Company.

Berry, A., & Browne, J. (2022). Mendel and darwin. *Proceedings of the National Academy of Sciences, 119*(30), e2122144119. https://doi.org/10.1073/pnas.2122144119

Bornmann, L., Tekles, A., Zhang, H. H., & Ye, F. Y. (2019). Do we measure novelty when we analyze unusual combinations of cited references? A validation study of bibliometric novelty indicators based on F1000Prime data. *Journal of Informetrics, 13*(4). https://doi.org/10.1016/j.joi.2019.100979

Bu, Y., Waltman, L., & Huang, Y. (2021). A multidimensional framework for characterizing the citation impact of scientific publications. *Quantitative Science Studies, 2*(1), 155-183. https://doi.org/10.1162/qss_a_00109

Chen, L., & Fang, H. (2019). An Automatic Method for Extracting Innovative Ideas Based on the Scopus® Database. *Knowledge Organization, 46*(3), 171-186. https://doi.org/10.5771/0943-7444-2019-3-171

Chen, Z., Zhang, C., Zhang, H., Zhao, Y., Yang, C., & Yang, Y. (2024). Exploring the relationship between team institutional composition and novelty in academic papers based on fine-grained knowledge entities. *The Electronic Library, ahead-of-print*(ahead-of-print). https://doi.org/10.1108/EL-03-2024-0070

Christensen, C. M., McDonald, R., Altman, E. J., & Palmer, J. E. (2018). Disruptive Innovation: An Intellectual History and Directions for Future Research. *Journal of Management Studies, 55*(7), 1043-1078. https://doi.org/10.1111/joms.12349

Cohen, B. A. (2017). How should novelty be valued in science? *elife, 6*. https://doi.org/10.7554/eLife.28699

Dirk, L. (1999). A measure of originality: the elements of Science. *Social Studies of Science, 29*(5),

756-777.

Fontana, M., Iori, M., Montobbio, F., & Sinatra, R. (2020). New and atypical combinations: An assessment of novelty and interdisciplinarity. *Research Policy, 49*(7). https://doi.org/10.1016/j.respol.2020.104063

Foster, J. G., Rzhetsky, A., & Evans, J. A. (2015). Tradition and Innovation in Scientists' Research Strategies. *American Sociological Review, 80*(5), 875-908. https://doi.org/10.1177/0003122415601618

Foster, J. G., Shi, F., & Evans, J. A. (2021). Surprise! Measuring novelty as expectation violation. *SocArXiv preprint*. https://doi.org/10.31235/osf.io/2t46f

Funk, R. J., & Owen-Smith, J. (2017). A dynamic network measure of technological change. *Management Science, 63*(3), 791–817. https://doi.org/10.1287/mnsc.2015.2366

Gates, A. J., & Barabási, A.-L. (2023). Reproducible science of science at scale: pySciSci. *Quantitative Science Studies*, 1-11. https://doi.org/10.1162/qss_a_00260

Guetzkow, J., Lamont, M., & Mallard, G. (2004). What is Originality in the Humanities and the Social Sciences? *American Sociological Review, 69*(2), 190-212.

Hargadon, A., & Sutton, R. I. (1997). Technology Brokering and Innovation in a Product Development Firm. *Administrative Science Quarterly, 42*(4), 716-749.

Hofstra, B., Kulkarni, V. V., Galvez, S. M.-N., He, B., Jurafsky, D., & McFarland, D. A. (2020). The Diversity–Innovation Paradox in Science. *Proc Natl Acad Sci U S A, 117*(17), 9284–9291. https://doi.org/10.1073/pnas.1915378117

Hou, J., Wang, D., & Li, J. (2022). A new method for measuring the originality of academic articles based on knowledge units in semantic networks. *Journal of Informetrics, 16*(3), 101306. https://doi.org/10.1016/j.joi.2022.101306

Janssen, M., Stoopendaal, A. M. V., & Putters, K. (2015). Situated novelty: Introducing a process perspective on the study of innovation. *Research Policy, 44*(10), 1974-1984. https://doi.org/10.1016/j.respol.2015.06.008

Jeon, D., Lee, J., Ahn, J. M., & Lee, C. (2023). Measuring the novelty of scientific publications: A fastText and local outlier factor approach. *Journal of Informetrics, 17*(4). https://doi.org/10.1016/j.joi.2023.101450

Jones, B. F. (2009). The burden of knowledge and the "death of the Renaissance man": Is innovation getting harder? *The Review of Economic Studies, 76*(1), 283–317.

Kaplan, N. (1965). The norms of citation behavior: Prolegomena to the footnote. *American Documentation, 16*(3), 179-184.

Ke, Q. (2020). Technological impact of biomedical research: The role of basicness and novelty. *Research Policy, 49*(7), 104071. https://doi.org/10.1016/j.respol.2020.104071

Ke, Q., Ferrara, E., Radicchi, F., & Flammini, A. (2015). Defining and identifying Sleeping Beauties in science. *Proc Natl Acad Sci U S A, 112*(24), 7426-7431. https://doi.org/10.1073/pnas.1424329112

Kuhn, T. S. (1977). *The structure of scientific revolutions*: University of Chicago press Chicago.

Lakatos, I. (1968). Criticism and the Methodology of Scientific Research Programmes. *Proceedings of the Aristotelian Society, 69*, 149-186.

Leahey, E., Lee, J., & Funk, R. J. (2023). What Types of Novelty Are Most Disruptive? *American Sociological Review, 88*(3), 562-597. https://doi.org/10.1177/00031224231168074

Lee, Y.-N., Walsh, J. P., & Wang, J. (2015). Creativity in scientific teams: Unpacking novelty and

impact. *Research Policy, 44*(3), 684–697. https://doi.org/10.1016/j.respol.2014.10.007

Lin, Y., Evans, J. A., & Wu, L. (2022). New directions in science emerge from disconnection and discord. *Journal of Informetrics, 16*, 101234. https://doi.org/10.1016/j.joi.2021.101234

Lin, Z., Yin, Y., Liu, L., & Wang, D. (2023). SciSciNet: A large-scale open data lake for the science of science research. *Sci Data, 10*(1), 315. https://doi.org/10.1038/s41597-023-02198-9

Liu, M., Bu, Y., Chen, C., Xu, J., Li, D., Leng, Y., et al. (2022). Pandemics are catalysts of scientific novelty: Evidence from COVID-19. *Journal of the Association for Information Science and Technology, 73*(8), 1065-1078. https://doi.org/10.1002/asi.24612

Luo, Z., Lu, W., He, J., & Wang, Y. (2022). Combination of research questions and methods: A new measurement of scientific novelty. *Journal of Informetrics, 16*(2). https://doi.org/10.1016/j.joi.2022.101282

Martín de Diego, I., González-Fernández, C., Fernández-Isabel, A., Fernández, R. R., & Cabezas, J. (2021). System for evaluating the reliability and novelty of medical scientific papers. *Journal of Informetrics, 15*(4). https://doi.org/10.1016/j.joi.2021.101188

Matsumoto, K., Shibayama, S., Kang, B., & Igami, M. (2021). Introducing a novelty indicator for scientific research: validating the knowledge-based combinatorial approach. *Scientometrics, 126*(8), 6891-6915. https://doi.org/10.1007/s11192-021-04049-z

Min, C., Bu, Y., Wu, D., Ding, Y., & Zhang, Y. (2021). Identifying citation patterns of scientific breakthroughs: A perspective of dynamic citation process. *Information Processing & Management, 58*(1). https://doi.org/10.1016/j.ipm.2020.102428

Mirowski, P. (2018). The future(s) of open science. *Soc Stud Sci, 48*(2), 171-203. https://doi.org/10.1177/0306312718772086

Mishra, S., & Torvik, V. I. (2016). Quantifying Conceptual Novelty in the Biomedical Literature. *Dlib Mag, 22*(9-10), 1-14. https://doi.org/10.1045/september2016-mishra

Morgan, P. P. (1985). Originality, novelty and priority: Three words to reckon with in scientific publishing. *Canadian Medical Association Journal, 132*(1), 8-9.

Mulkay, M. (1974). Conceptual Displacement and Migration in Science: A Prefatory Paper. *Science Studies, 4*(3), 205-234.

Nelson, R. R., & Winter, S. G. (1982). *An evolutionary theory of economic change*: Belknap Press of Harvard University Press.

Packalen, M., & Bhattacharya, J. (2019). Age and the trying out of new ideas. *Journal of Human Capital, 13*(2), 341-373. https://doi.org/10.1086/703160

Park, M., Leahey, E., & Funk, R. J. (2023). Papers and patents are becoming less disruptive over time. *Nature, 613*(7942), 138–144. https://doi.org/10.1038/s41586-022-05543-x

Pelletier;, P., & Wirtz;, K. (2022). Novelpy: A Python package to measure novelty and disruptiveness of bibliometric and patent data. *arxiv*.

Porter, A. L., & Rafols, I. (2009). Is science becoming more interdisciplinary? Measuring and mapping six research fields over time. *Scientometrics, 81*(3), 719-745. https://doi.org/10.1007/s11192-008-2197-2

Raan, A. F. J. V. (2004). Sleeping Beauties in science. *Scientometrics, 59*(3), 467-472.

Rodríguez-Navarro, A. (2015). Research assessment based on infrequent achievements: A comparison of the United States and Europe in terms of highly cited papers and Nobel Prizes. *Journal of the Association for Information Science and Technology, 67*(3), 731-740. https://doi.org/10.1002/asi.23412

Ruan, X., Ao, W., Lyu, D., Cheng, Y., & Li, J. (2023). Effect of the topic-combination novelty on the disruption and impact of scientific articles: Evidence from PubMed. *Journal of Information Science*. https://doi.org/10.1177/01655515231161133

Schumpeter, J. A. (1934). *The Theory of Economic Development*. Harvard University Press, Cambridge Mass.

Shi, F., & Evans, J. (2023). Surprising combinations of research contents and contexts are related to impact and emerge with scientific outsiders from distant disciplines. *Nat Commun, 14*(1), 1641. https://doi.org/10.1038/s41467-023-36741-4

Shibayama, S., & Wang, J. (2019). Measuring originality in science. *Scientometrics, 122*(1), 409-427. https://doi.org/10.1007/s11192-019-03263-0

Shibayama, S., Yin, D., & Matsumoto, K. (2021). Measuring novelty in science with word embedding. *PLoS One, 16*(7), e0254034. https://doi.org/10.1371/journal.pone.0254034

Silera, K., Lee, K., & Bero, L. (2014). Measuring the effectiveness of scientific gatekeeping. *Proc Natl Acad Sci U S A, 112*(2), 360-365. https://doi.org/10.1073/pnas.1418218112

Sinatra, R., Wang, D., Deville, P., Song, C., & Barabasi, A. L. (2016). Quantifying the evolution of individual scientific impact. *Science, 354*(6312). https://doi.org/10.1126/science.aaf5239

Small, H. (1973). Co-citation in the scientific literature: A new measure of the relationship between two documents. *Journal of the American Society for Information Science, 24*(4), 265-269. https://doi.org/10.1002/asi.4630240406

Tahamtan, I., & Bornmann, L. (2018). Creativity in science and the link to cited references: Is the creative potential of papers reflected in their cited references? *Journal of Informetrics, 12*(3), 906-930. https://doi.org/10.1016/j.joi.2018.07.005

Thelwall, M. (2024). Can ChatGPT evaluate research quality? *Journal of Data and Information Science, 9*(2), 1-21. https://doi.org/10.2478/jdis-2024-0013

Thelwall, M., & Sud, P. (2022). Scopus 1900–2020: Growth in articles, abstracts, countries, fields, and journals. *Quantitative Science Studies, 3*(1), 37-50. https://doi.org/10.1162/qss_a_00177

Trapido, D. (2015). How novelty in knowledge earns recognition: The role of consistent identities. *Research Policy, 44*(8), 1488-1500. https://doi.org/10.1016/j.respol.2015.05.007

Trumble, W., Siefring, J., & Bailey, C. (2007). The shorter Oxford English dictionary (6th ed.). New York, NY: Oxford University Press.

Uzzi, B., Mukherjee, S., Stringer, M., & Jones, B. (2013). Atypical combinations and scientific impact. *Science, 342*(6157), 468-472. https://doi.org/10.1126/science.1240474

Verhoeven, D., Bakker, J., & Veugelers, R. (2016). Measuring technological novelty with patent-based indicators. *Research Policy, 45*(3), 707-723. https://doi.org/10.1016/j.respol.2015.11.010

Veugelers, R., & Wang, J. (2019). Scientific novelty and technological impact. *Research Policy, 48*, 1362-1372. https://doi.org/10.1016/j.respol.2019.01.019

Waltman, L. (2016). A review of the literature on citation impact indicators. *Journal of Informetrics, 10*(2), 365-391. https://doi.org/10.1016/j.joi.2016.02.007

Wang, J., Veugelers, R., & Stephan, P. (2017). Bias against novelty in science: A cautionary tale for users of bibliometric indicators. *Research Policy, 46*(8), 1416-1436. https://doi.org/10.1016/j.respol.2017.06.006

Wang, S., Ma, Y., Mao, J., Bai, Y., Liang, Z., & Li, G. (2022). Quantifying scientific breakthroughs by a novel disruption indicator based on knowledge entities. *Journal of the Association for Information Science and Technology, 74*(2), 150-167. https://doi.org/10.1002/asi.24719

Wang, Z., Zhang, H., Chen, J., & Chen, H. (2024). Measuring the novelty of scientific literature through contribution sentence analysis using deep learning and cloud model. *SSRN*.

Wu, L., Wang, D., & Evans, J. A. (2019). Large teams develop and small teams disrupt science and technology. *Nature, 566*(7744), 378–382. https://doi.org/10.1038/s41586-019-0941-9

Wuestman, M., Hoekman, J., & Frenken, K. (2020). A typology of scientific breakthroughs. *Quantitative Science Studies, 1*(3), 1203-1222. https://doi.org/10.1162/qss_a_00079

Yan, Y., Tian, S., & Zhang, J. (2019). The impact of a paper's new combinations and new components on its citation. *Scientometrics, 122*(2), 895-913. https://doi.org/10.1007/s11192-019-03314-6

Zhao, Y., Liu, L., & Zhang, C. (2022). Is coronavirus-related research becoming more interdisciplinary? A perspective of co-occurrence analysis and diversity measure of scientific articles. *Technological Forecasting and Social Change, 175*, 121344. https://doi.org/10.1016/j.techfore.2021.121344

Zhu, N., Zhang, D., Wang, W., Li, X., Yang, B., Song, J., et al. (2020). A Novel Coronavirus from Patients with Pneumonia in China, 2019. *N Engl J Med, 382*(8), 727-733. https://doi.org/10.1056/NEJMoa2001017

# Appendix A.

Table A.1 the list of literature for the review

| ID | Title | Journal | Year |
|---|---|---|---|
| 1 | An effective framework for measuring the novelty of scientific articles through integrated topic modeling and cloud model | *Journal of Informetrics* | 2024 |
| 2 | Exploring the relationship between team institutional composition and novelty in academic papers based on fine-grained knowledge entities | *The Electronic Library* | 2024 |
| 3 | Reproducible science of science at scale: pySciSci | *Quantitative Science Studies* | 2023 |
| 4 | Measuring the novelty of scientific publications: A fastText and local outlier factor approach | *Journal of Informetrics* | 2023 |
| 5 | What Types of Novelty Are Most Disruptive? | *American Sociological Review* | 2023 |
| 6 | SciSciNet: A large-scale open data lake for the science of science research | *Scientific Data* | 2023 |
| 7 | Effect of the topic-combination novelty on the disruption and impact of scientific articles: Evidence from PubMed. | *Journal of Information Science*. | 2023 |
| 8 | Surprising combinations of research contents and | *Nature* | 2023 |

|  | contexts are related to impact and emerge with scientific outsiders from distant disciplines. | *Communications* |  |
|---|---|---|---|
| 9 | A new method for measuring the originality of academic articles based on knowledge units in semantic networks | *Journal of Informetrics* | 2022 |
| 10 | New directions in science emerge from disconnection and discord | *Journal of Informetrics* | 2022 |
| 11 | Pandemics are catalysts of scientific novelty: Evidence from COVID-19. | *Journal of the Association for Information Science and Technology* | 2022 |
| 12 | Combination of research questions and methods: A new measurement of scientific novelty. | *Journal of Informetrics* | 2022 |
| 13 | Novelpy: A Python package to measure novelty and disruptiveness of bibliometric and patent data | *Arxiv* | 2022 |
| 14 | Surprise! Measuring novelty as expectation violation | *SocArXiv preprint* | 2021 |
| 15 | System for evaluating the reliability and novelty of medical scientific papers | *Journal of Informetrics* | 2021 |
| 16 | Introducing a novelty indicator for scientific research: validating the knowledge-based combinatorial approach. | *Scientometrics* | 2021 |
| 17 | Measuring novelty in science with word embedding | *plos one* | 2021 |
| 18 | The Diversity–Innovation Paradox in Science | *Proc Natl Acad Sci U S A* | 2020 |
| 19 | Technological impact of biomedical research: The role of basicness and novelty | *Research Policy* | 2020 |
| 20 | New and atypical combinations: An assessment of novelty and interdisciplinarity | *Research Policy* | 2020 |
| 21 | Do we measure novelty when we analyze unusual combinations of cited references? A validation study of bibliometric novelty indicators based on F1000Prime data | Journal of Informetrics | 2019 |
| 22 | An Automatic Method for Extracting Innovative Ideas Based on the Scopus® Database | *Knowledge Organization,* | 2019 |
| 23 | Age and the Trying Out of New Ideas | *Journal of Human Capital* | 2019 |
| 24 | The impact of a paper's new combinations and new components on its citation | *Scientometrics* | 2019 |
| 25 | Network-based approach to detect novelty of scholarly literature. | Information Sciences | 2018 |
| 26 | Creativity in science and the link to cited references: Is the creative potential of papers reflected in their cited references? | *Journal of Informetrics* | 2018 |

| 27 | Bias against novelty in science: A cautionary tale for users of bibliometric indicators | *Research Policy* | 2017 |
| 28 | Looking Across and Looking Beyond the Knowledge Frontier: Intellectual Distance, Novelty, and Resource Allocation in Science | *Management Science* | 2016 |
| 29 | Quantifying Conceptual Novelty in the Biomedical Literature | *D-Lib Magazine* | 2016 |
| 30 | Tradition and Innovation in Scientists' Research Strategies | *American Sociological Review* | 2015 |
| 31 | Situated novelty: Introducing a process perspective on the study of innovation | *Research Policy* | 2015 |
| 32 | Creativity in scientific teams: Unpacking novelty and impact | *Research Policy* | 2015 |
| 33 | How novelty in knowledge earns recognition: The role of consistent identities. | *Research Policy* | 2015 |
| 34 | Atypical combinations and scientific impact | *Science* | 2013 |
| 35 | Incentives and creativity: evidence from the academic life sciences | *the RAND Journal of Economics* | 2011 |
| 36 | What is Originality in the Humanities and the Social Sciences? | *American Sociological Review* | 2004 |
| 37 | Originality, novelty and priority: Three words to reckon with in scientific publishing | *Canadian Medical Association Journal* | 1985 |